\documentclass[11pt]{article}

\usepackage{graphicx}

\begin{document}
 \bibliographystyle{unsrt}
\title{\Large \bf Electron and Ion Transport in Dense Rare Gases\footnote{paper submitted to IEEE--TDEI}}
\author{{\large \bf A. F. Borghesani}\\
Department of Physics \\
University of Padua\\
via F. Marzolo,8 \\
I--35131 Padua, Italy}

\maketitle

\begin{abstract}
A review of the research on electron and ion transport in dense rare gases is presented. The investigation of the transport properties of electrons in dense rare gases aims at understanding the dynamics and energetics of electron states in a dense medium and at elucidating how changes of the environment influence their nature and scattering properties. The quantum nature of electrons couples them to the environment is such a way to produce a density-dependent shift of their energy that is the key to rationalize the observed phenomena.
\end{abstract}

P.A.C.S.: 51.50.+v, 52.25.Fi, 51.10.+y, 71.20.-h, 71.50.+t, 33.20.Ea, 33.70.Jg, 34.50.Gb
\section{Introduction}
The conduction of excess electrons in non polar dielectric liquids is a topic of practical and fundamental relevance. For example, the drift of electrons in liquefied noble gases is the working principle of ionizing radiation detectors, which are routinely used in high-energy physics. Thus, the design of many industrial and research apparatuses requires the knowledge of the transport properties of excess electrons in dielectric media. Significant pieces of information on the fundamental electron-atom scattering processes in a dense medium is gathered by investigating electron transport carefully \cite{schmidt97,schmidt89}.

In particular, liquid rare gases are the simplest possible systems. Their atoms are approximately spherically symmetric and have quite large ionization energies so that ionization processes do not contribute significantly at the small energies typically involved in transport experiments. Rare gases are extremely well characterized: the electron-atom cross sections are well known \cite{zecca} as are their equations of state. Finally, most important from the experimental point of view, samples of compressed or liquefied noble gases of extremely high purity can be quite easily produced.

Typically, swarm techniques are used to study electrons transport \cite{schmidt97}. In the simplest experiments, electrons injected into the liquid under investigation are drifted through it by applying a suitable uniform electric field E across two plane and parallel electrodes. The analysis of the signal induced at the collector allows to determine the drift time and hence the drift velocity $v_D.$ According to classic kinetic theory \cite{huxley}, it is customary to calculate the electron mobility 
$\mu =v_{D}E$ because it is directly related to the scattering cross section.

As an example, the electron mobility $\mu$ measured in liquid Ar at the normal boiling point for $T=87.3$ K and $P=0.1$ MPa, is shown in Figure 1 as a function of the externally applied electric field $E.$ 
\begin{figure}\begin{center}
\includegraphics[scale=0.4]{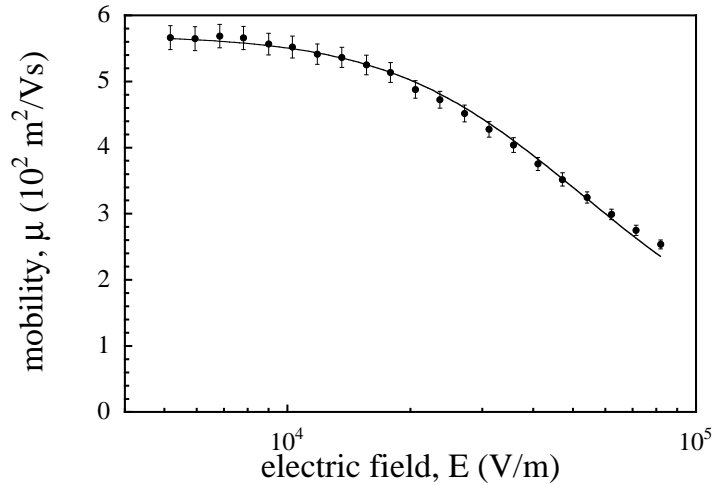}
\caption{\small  Electron mobility in liquid argon at the normal boiling point as a function of the electric field \cite{bic}.}\end{center}
    \label{figure1}
\end{figure}
It is interesting to note that the field dependence of the electron  mobility in the cryogenic liquid is similar to that of an electron in a dilute hard--sphere gas. For this reason, a gas--kinetic approach has been adopted to describe the experimental results \cite{bic,kaneko88}.

According to standard kinetic theory \cite{huxley,cohen} $\mu$ is related to the scattering cross section by the following relationship
\begin{equation}
\mu=-\left( 
\frac{e}{3}\right)\left( 
\frac{2}{m}\right)^{1/2}\int\limits_0^\infty \frac{\epsilon}{N\sigma_{mt}\left(\epsilon\right)}\left[ \frac{\mathrm{d}g
\left(\epsilon\right)}{\mathrm{d}\epsilon}\right]\mathrm{d}\epsilon
\label{eq:1}
\end{equation}
where $\sigma_{mt}(\epsilon)$ is the energy-dependent momentum-transfer cross section of the electron--atom interaction, $m$ and $e$ are the electron mass and charge, respectively. $N$ is the gas number density. $ \epsilon$ is the electron energy and $g(\epsilon )$ is the Davydov-Pidduck distribution function \cite{huxley,cohen} given by
\begin{equation}
\ln{g\left(\epsilon\right)/A}  = -\int\limits_0^\epsilon
\frac{\mathrm{d}z}{k_{\mathrm{B}T} +\left(\frac{M}{6m}\right)
\left( \frac{E}{N}\right)^2 \left[z\sigma_{mt}(z)\sigma_E(z)\right]^{-1}
}
\label{eq:2}
\end{equation}
$A$ is a normalization constant. The distribution function is normalized in such a way that
\begin{equation}
\int\limits_0^\infty z^{1/2} g(z) \mathrm{d} z =1
\label{eq:3}
\end{equation}
Finally, $\sigma_E (\epsilon)$ is the energy-transfer cross section.

An accurate description of the experimental data is obtained by assuming $\sigma_{mt} (\epsilon)=\sigma_{m0}=$ constant and $\sigma_{E} (\epsilon)=\sigma_{E0}\pi k_\mathrm{B} T/\epsilon,$ where $k_\mathrm{B}$ is the Boltzmann constant. $\sigma_{m0}$ and $\sigma_{E0}$ are adjusted so as to fit the data as shown in Figure 1, in which the solid line is obtained by letting $\sigma_{m0}=0.213$ \AA$^2$ and $\sigma_{E0}=10$ \AA$^2$ \cite{bic}.

The agreement of the gas-kinetic model with the experiment in a liquid raises the question of how the solution of the Boltzmann transport equation can be successfully used in a dense environment.

At first glance, one immediately realizes that the cross sections in equations \ref{eq:1} and \ref{eq:2} are adjustable parameters, whereas the cross sections in dilute gas systems are given either theoretically from an electron-atom interaction potential or experimentally from swarm- or beam experiments \cite{zecca}.

It is clear that, as the gas density is increased from dilute values, the interatomic distance decreases and the quantum nature of the electron cannot be neglected anymore, especially at low temperature. The electron wave packet interacts simultaneously with many atoms at once and several multiple scattering effects come into play. 

The effects of the environmental changes can be summarized by introducing a density-dependent shift of the ground state electron energy and a density-dependent modification of the cross sections that progressively go away from their gas-phase values to rather become effective cross sections \cite{bsl}.

The investigation of the electron dynamics and energetics of electrons in dense gases as a function of the gas density is thus aimed at following the transition from true- to effective-cross sections that allows to retain the gas-kinetic picture even in the liquid.

This paper summarizes the experimental results on electron transport in dense noble gases and on some other properties of electrons in such an environment, which yield a consistent picture of the states of excess electrons in a dense medium.

\section{Transport properties}\label{sec:trnsprt}
The attention of this paper is focused on dense rare gases because they are the simplest systems. Their atoms are spherical symmetric, their interaction with electrons is well known and there is a plenty of information about their cross sections \cite{zecca}. These facts make rare gases the best systems to elucidate the basic physical mechanisms that determine transport.

In the limit of vanishingly small electric field,$ E=0,$ kinetic theory predicts that the density-normalized mobility m0N of electrons in a gas at constant temperature T is independent of the density N \cite{huxley}. This result as can  easily obtained by carrying out the integrations in equations \ref{eq:1} through \ref{eq:3}. 
\begin{figure}\begin{center}
\includegraphics[scale=0.4]{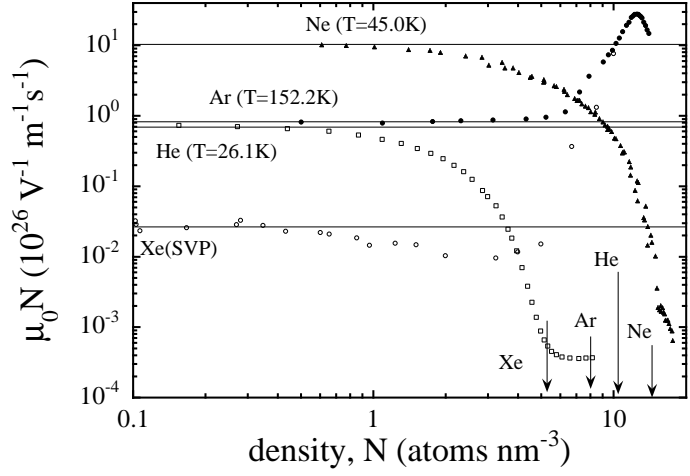}
\caption{\small Zero-field density-normalized mobility$\mu_0 N$ as a function of the gase density $N$ for He (squares) \cite{borg02}, Ne (triangles) \cite{borg90}, Ar (closed dots) \cite{cpl,borg01}, Xe (open dots) \cite{huang}. The constant lines are the prediction of kinetic theory. Arrows indicate the respective critical densities.}\end{center}
    \label{figure2}
\end{figure}

In Figure 2 experimental results obtained for the zero-field density-normalized mobility $\mu_0 N$ in He, Ne, Ar, and Xe are shown \cite{bsl,borg02,borg90,huang,borg88}. Constant lines in the picture indicate the predictions of kinetic theory that are strongly violated in real experiments. 

The reduced mobility in He and Ne decreases by more than 4 orders of magnitude as $N$ is increased up to values comparable or larger than the liquid density at the critical point. On the contrary, $\mu_0 N$ in Ar increases up to a value nearly 30 times larger than the low-density value, whereas Xe shows a feeble drop at first and then a nearly 3 orders-of-magnitude increase with increasing $N$ before its mobility drops again at even larger density.

Whereas the mobility measurements in He \cite{borg02}, Ne \cite{borg88,borg90}, and Ar \cite{cpl,borg01} have been carried out along isotherms, the Xe results have been obtained at saturated vapor pressure \cite{huang}. In this way, the temperature $T$ is not constant for all $N$ and the analysis is more complicated than for He, Ne, and Ar, although the conclusions are the same. 

In a more expanded picture (see Figure 3), deviations of the data from the predictions of kinetic theory appear even at the lowest densities.
\begin{figure}\begin{center}
\includegraphics[scale=0.4]{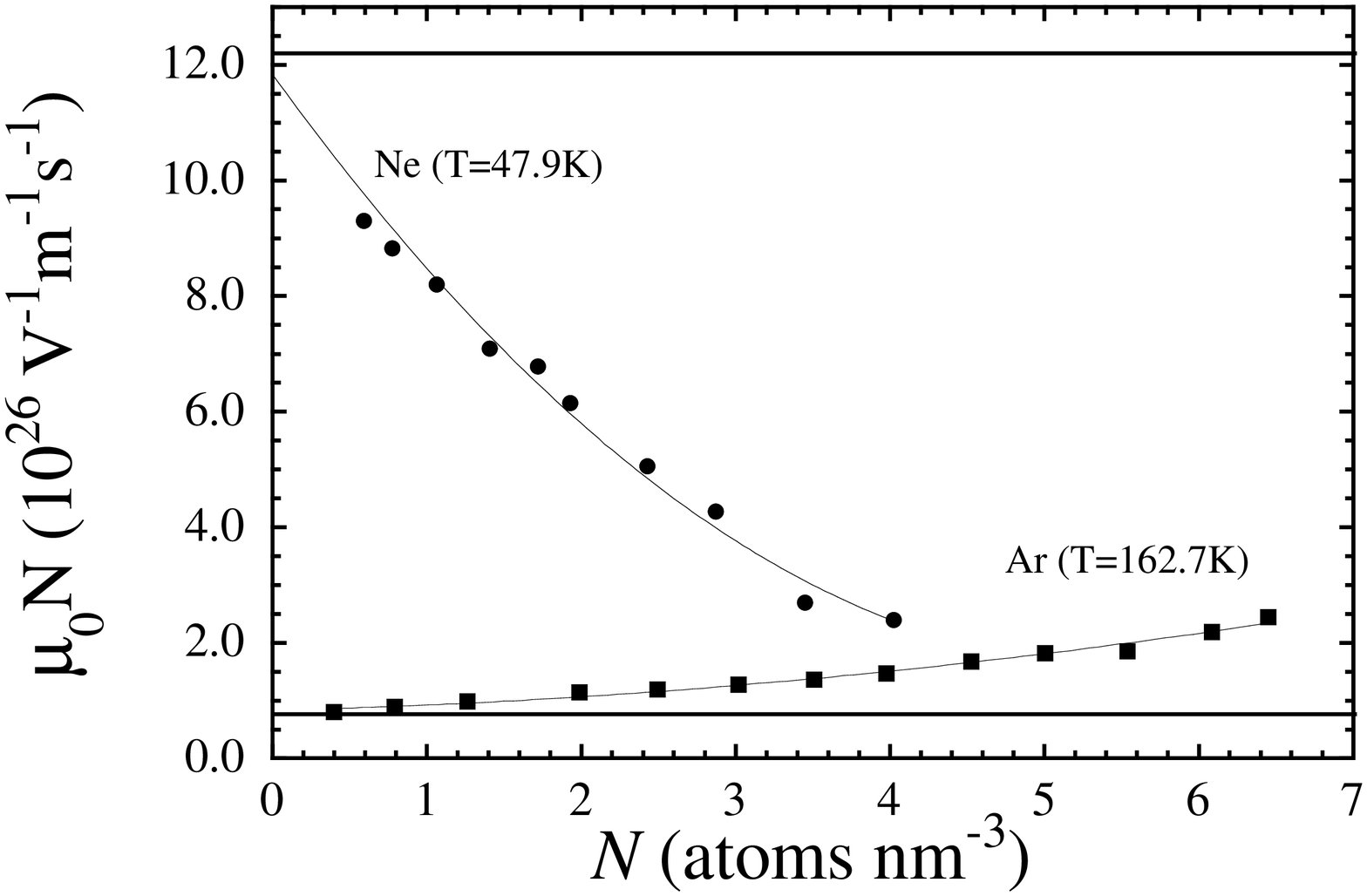}
\caption{\small Zero-field density-normalized mobility $ \mu_0 N$ as a function of $N$ for Ne at $T=47.9$ K (dots) \cite{borg88} and  for Ar at $T=162.7$ K (squares) \cite{cpl}. Constant lines: kinetic theory. Lines through the data: heuristic model.}\end{center}
    \label{figure3}
\end{figure}
 The positive (Ar \cite{cpl}) and negative (Ne \cite{borg88}) deviations from the kinetic predictions are shown for low-- to intermediate $N.$

As a general rule, the mobility in gases of small polarizability (He and Ne) shows negative deviations. More polarizable gases (Ar and the heavier noble gases) show positive deviations.

The electron-atom interaction in He and Ne is actually dominated by short-range repulsive exchange forces and is characterized by a positive scattering length. On the contrary, the interaction between low-energy excess electrons and the atoms of the heavier noble gases is mainly due to long-range polarization forces and the scattering length is negative \cite{zecca}.

Ne \cite{borg88}] and He \cite{borg02,levi} show a further dramatic decrease of the mobility at high density until $\mu_0 N$ levels off at very small values, as shown in Figure 2. This drop is commonly associated to self--trapping of electrons in (partially) empty cavities, known as bubbles \cite{hern}, endowed with very small hydrodynamic mobility, whose existence has been reported for the first time in superfluid He \cite{kuper}. The smooth transition to low mobility values observed in Figure 2 depends on the fact that the observed mobility is an average over extended, highly mobile-- and localized, low--mobility states. 

The behavior of the transport properties of electrons must be thus analyzed in a first region up to intermediate $N$ values, in which the electron transport proceeds via extended states, and in a second region for higher $N,$ in which the coupling between electron and environment leads to the formation of a new kind of states. 

\subsection{Transport up to intermediate density}
The basic assumption of kinetic theory is that the density of scatterers  is so low that only binary collisions take place. The electron mean free path $\ell ,$ the atomic size $d,$ the average interatomic distance $N^{-1/3},$ and the electron thermal wavelength $\lambda_T$  must satisfy the inequalities   $\ell \gg N^{-1/3} > d $ and  $\ell \gg \lambda_T.$ In this case, the two-terms solution of the Boltzmann transport equation yields equations (1) and (2) \cite{huxley,cohen}. However, if the last condition is not met, multiple scattering effects come into play \cite{lax,borg94}.

Several theories have been developed in order to explain the positive and negative density effects in terms of the sign of the scattering length. However, a unified picture of the scattering mechanisms in a dense gas can be heuristically developed \cite{borg94}.

It is known that electrons injected into liquid He must overcome a 1--eV--high barrier \cite{bro}. This means that, in a liquid, the average electron energy is different from (3/2)$k_{\mathrm{B}}T.$ It is also known that the spectral lines of alkali atoms in a buffer noble gas show a density-dependent shift, which is interpreted in terms of multiple scattering \cite{fermi}.

According to Springett, Cohen, and Jortner \cite{scj} the energy spectrum of an excess electron in a monoatomic fluid takes the form 
\begin{equation}
\epsilon^\prime = \frac{\hbar^2 k^2}{2m}+ V_0 (N)
\label{eq:4}
\end{equation}
which expresses the fact that the bottom of the conduction band is shifted with respect to {\em vacuum} by the density-dependent contribution
\begin{equation}
V_0 (N) = E_K (N) +U_P (N)
\label{eq:5}
\end{equation}
$U_P (N)$ is a potential energy term that takes into account the effect of the polarization of the medium. $E_K (N)$ is a kinetic energy term related to the quantum nature of the electron. It can be calculated by assuming that in the fluid there is a short-range order on the scale of the Wigner-Seitz radius $r_s=(3/4N)^{1/3}$ and by imposing the the condition that the electron wave function is invariant under a translation of $2r_s:$ $\psi (r)= \psi (r+2rs).$ For low-energy electrons of wave vector $k_0,$ the following eigenvalue equation is obtained [22]
\begin{equation}
\tan{\left[ k_0\left(r_s -\tilde a \left( k_0 \right)\right) \right] }= k_0 r_s
\label{eq:6}
\end{equation}
$\tilde a = (\sigma_T /4\pi)^{1/2}$ is the positive scattering length of the hard-core repulsive electron-atom pseudopotential and has the meaning of an effective atomic radius [21]. $\sigma_T$ is the total cross section. The kinetic energy shift $E_K$ is obtained as  
\begin{equation}
E_K (N) = \frac{\hbar^2 k_0^2}{2m}
\label{eq:7}
\end{equation}
Only $E_K$ affects the dynamical properties of the electrons. The group velocity   contributes to the equipartition value arising from the gas temperature. The bottom of the energy distribution function is shifted by $E_K.$

This fact explains to a large extent the density dependence of the mobility $\mu 0,$ if the energy dependence of the momentum-transfer cross section of the different gases is considered (see Figure 4). 
\begin{figure}\begin{center}
\includegraphics[scale=0.4]{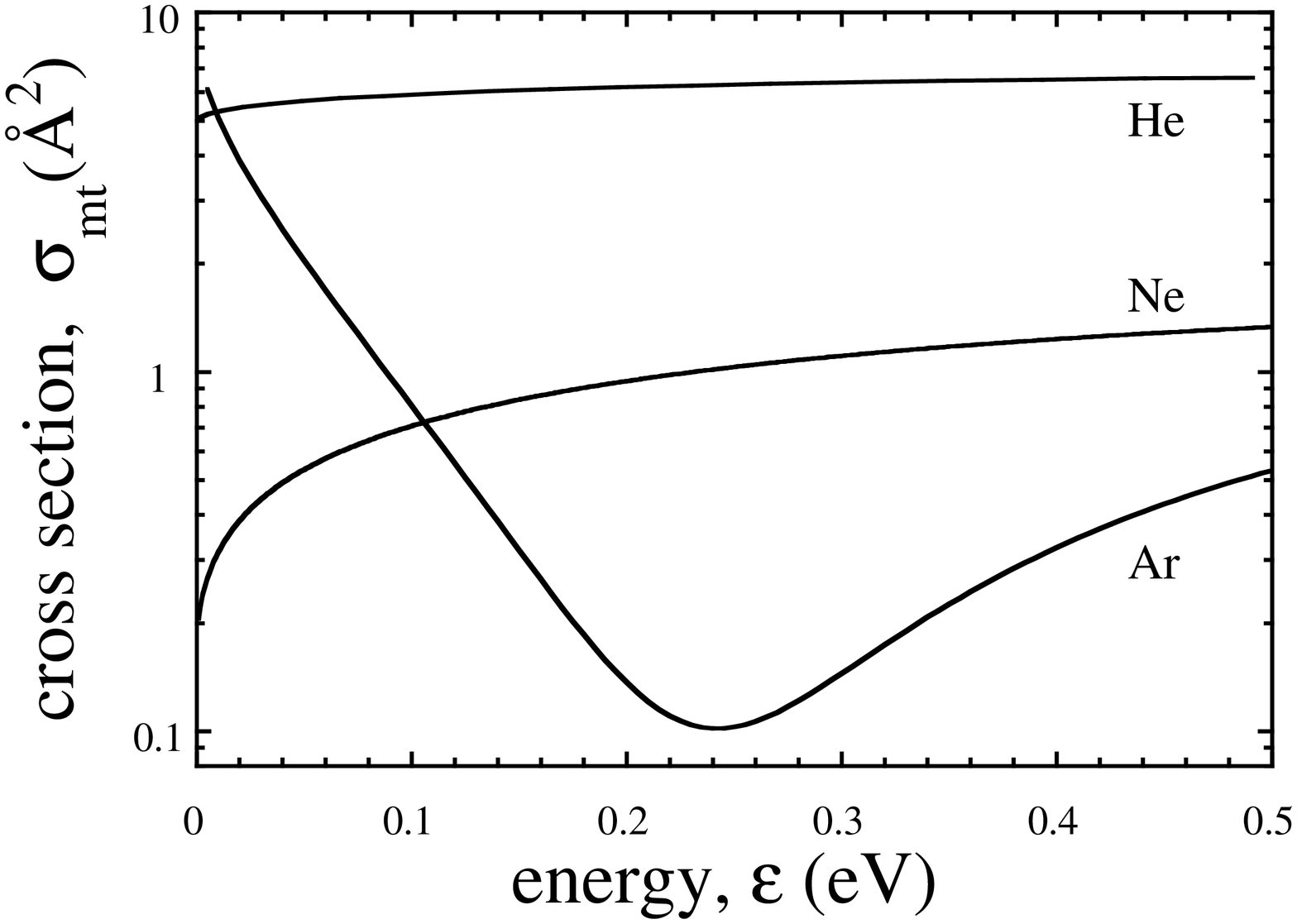}
\caption{\small  Energy dependence of the momentum transfer cross section for He, Ne, and Ar. Ar shows the Ramsauer-Townsend (RT) minimum \cite{zecca}.}\end{center}
    \label{figure4}
\end{figure}

In fact, $\mu_0 N$ is obtained by setting $E=0$ in equation \ref{eq:2}, yielding
\begin{equation}
\mu_0 N = \frac{4e}{3\sqrt{2\pi m\left( k_\mathrm{B}T\right)^5}}\int\limits_0^\infty
\frac{\epsilon}{\sigma_{mt}\left(\epsilon\right)}e^{-\epsilon/k_\mathrm{B}T}\mathrm{d}\epsilon
\label{eq:8}
\end{equation}
$\mu_0 N$ is a sort of average of the inverse momentum transfer cross section. If the distribution function is shifted by the positive, density-dependent contribution $E_K (N),$ the cross section is sampled at an increasingly higher mean energy. Thus, the mobility decreases if the cross section increases with the energy and, conversely, it decreases if the cross section increases with energy.

This effect is small for He, for which smt is nearly constant and is very large for Ne and Ar, whose cross sections depend strongly on the electron energy. In Ne, a positive $N-$dependent energy shift increases the average cross section, thereby reducing $\mu_0$ \cite{borg88}. The opposite effect occurs in Ar, as shown in Figure 3 \cite{bsl,borg01}.

The fact that the energy shift $E_K$ is positive can be deduced by the analysis of the field dependence of the mobility $\mu$ in Ar, for instance. In Figure 5 the density-normalized mobility $\mu N$ is shown as a function of the reduced electric field $E/N$ in Ar at $T=162.7$ K for several densities ($0.37<N<6.1$ atoms/nm$^3$) \cite{bsl}. 
\begin{figure}\begin{center}
\includegraphics[scale=0.4]{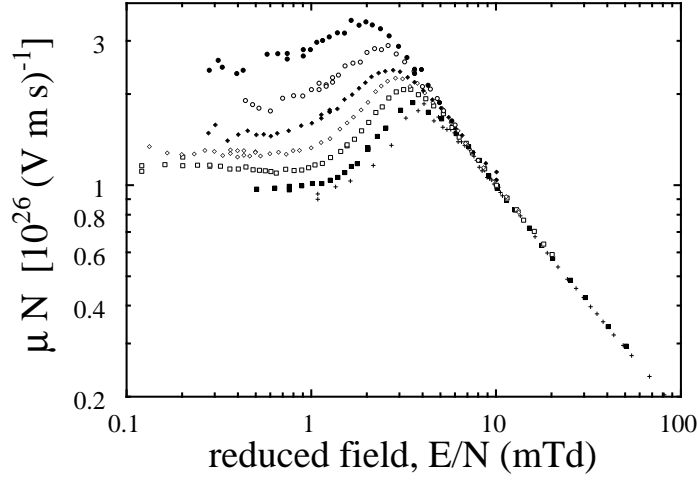}
\caption{\small  Density-normalized mobility $\mu N$ in Ar for several densities $N$ at $T=162.7$ K \cite{bsl}. 1 mTd = $10^{-24}$ Vm$^2.$ $N=$ 0.37, 0.791, 1.99, 3.02, 3.98, 5.0, and 6.08 atoms/nm$^3.$}\end{center}
    \label{figure5}
\end{figure}

In correspondence of the field $(E/N)_{max},$ at which the mobility is maximum for each $N,$ the average electron energy equals that of the RT--mi\-ni\-mum. $(E/N)_{max}$ decreases with increasing $N$ at constant $T,$ as shown in Figure 6.
\begin{figure}\begin{center}
\includegraphics[scale=0.4]{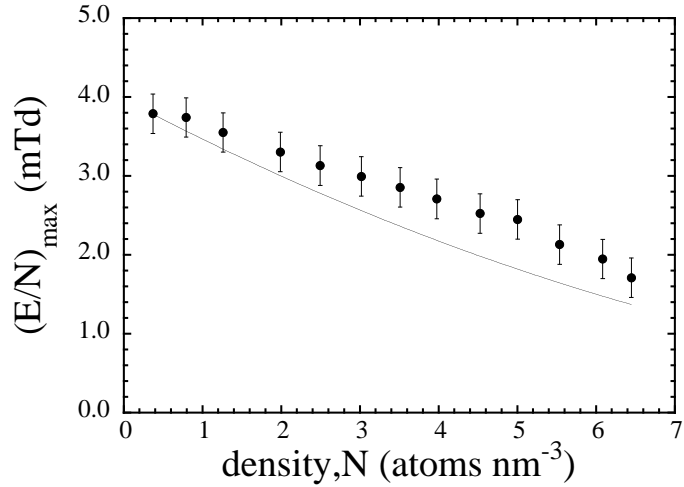}
\caption{\small  Reduced field$(E/N)_{max}$ of the mobility maximum in Ar as a function of $N$ for $T=162.7$ K \cite{bsl}. 1 mTd = 10$^{-24}$ Vm$^2.$ Solid line: heuristic model \cite{bsl}.}\end{center}
    \label{figure6}
\end{figure}

Because the quantity $(E/N)$ is proportional to the energy gained by electrons from the field in a mean free path, the conclusion is drawn that, upon increasing $N,$ electrons need a smaller contribution from the field in order that their average energy equals that of the RT minimum. This positive contribution is $E_K (N)$ \cite{bsl}.

In He, whose cross section is large but nearly constant, the previous effect is less important. In this case, another multiple scattering effect is more effective because the cross section is so large.

When $N$ increases, the electron mean free path decreases accordingly as $1/N\sigma$ and it may eventually become comparable to its thermal de Broglie wavelength. In this case, there is an enhancement of electron backscattering due to quantum self-interference of the electron wave function scattered off atoms located along paths that are connected by time-reversal symmetry. This phenomenon is closely related to the weak localization regime of the conduction in disordered solids and to the Anderson localization. It depends on the ratio of the electron wavelength to its mean free path $\lambda_T / \ell = N\sigma_{mt}\lambda_T$ and contributes most for large cross-section atoms.

For low $N,$ cross sections are increased by a factor $1+ N\sigma_{mt}\lambda/\pi$ \cite{atra}. At very high density, this factor leads to the appearance of a mobility edge \cite{poli}.

A final multiple scattering effect is related to the fact that, at high$ N,$ the electron wave function spans a large region including many atoms. The total scattered wave function is thus a coherent sum of partial contributions scattered off each individual atoms \cite{lekner}. The correlation among atoms, which is particularly strong near the critical point, yields a further enhancement of the cross section by a wave vector dependent factor $F(k),$ which can be accounted for once the static structure factor $S(k)$ is known
\begin{equation}
F(k)=\frac{1}{4k^4}\int\limits_0^{2k} q^3 S(q)\mathrm{d}q
\label{eq:9}
\end{equation}

The static structure factor, especially near the critical point, can be expressed as \cite{stanley}
\begin{equation}
S(q)= \frac{S(0)+(qL)^2}{1+(qL)^2}
\label{eq:10}
\end{equation}
where the long--range correlation length is given by $L^2=0.1l^2 [S(0)-1],$ and $l=10$ \AA\ is the short--range correlation length. $S(0)=Nk_\mathrm{B} T K_T$ is the long wavelength limit of the static structure factor  and $K_T$ is the isothermal compressibility.

The action of these three multiple scattering effects can be heuristically combined together to yield the effective cross section
\begin{equation}
\sigma_{mt}^\star \left( w\right)  = F\left( w \right) 
\sigma_{mt} \left( w \right)
\left[ 1 + \frac{2\hbar N F \left( w \right) \sigma_{mt} \left( w \right)}{\left( 2mw\right)^{1/2}}
\right]
\label{eq:11}
\end{equation}
where $w=\epsilon +E_K (N)$ is the shifted energy \cite{bsl, borg01,borg88}. The cross section is evaluated at the shifted energy and is enhanced by the weak localization- and correlation factors. The effective cross section is then inserted in equations \ref{eq:1} and \ref{eq:2} with the results shown in Figures  6, 7, and 8.
\begin{figure}\begin{center}
\includegraphics[scale=0.4]{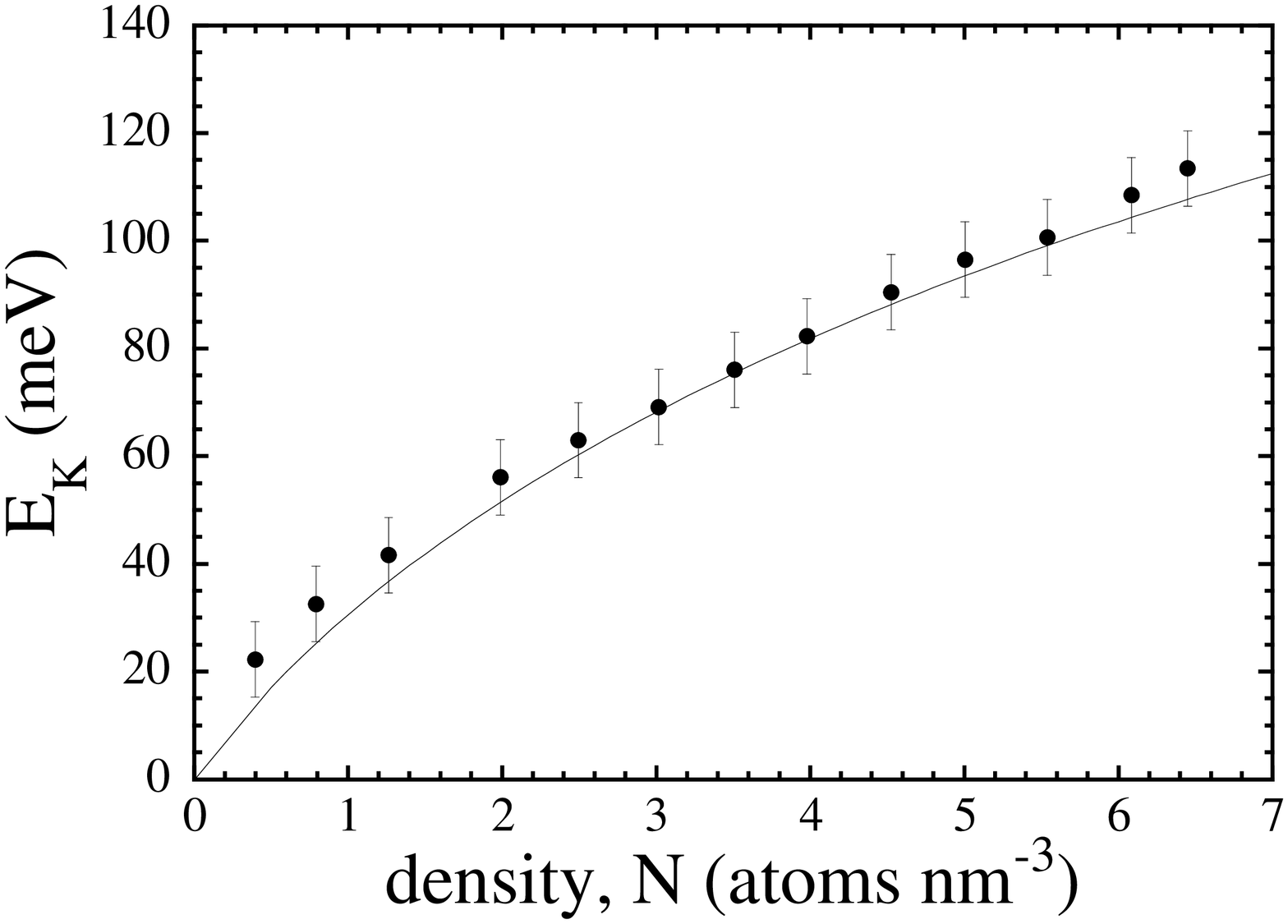}
\caption{\small  $E_K(N)$ in Ar for $T=162.7$ K \cite{bsl}. Solid line: Wigner-Seitz model.}\end{center}
    \label{figure7}
\end{figure}

In Figure 7 the values of$ _E$K to be inserted in equation \ref{eq:11} in order to obtain agreement between the heuristic model and the experimental determination of $\mu_0 N$ in Ar at $T=162.7$ K are compared with the predictions of equations \ref{eq:6} and \ref{eq:7} \cite{bsl}. The model appears to be highly consistent.

A further confirmation of the self-consistency of this heuristic model is its ability to predict with fair accuracy the density dependence of the quantity $(E/N)_{max},$ as shown in Figure 6 \cite{bsl}.

The model works equally well also for negative-density-effect gases. In Figure 8, the zero-field mobility$ \mu_0$ in Ne at $T=45$ K is compared with the predictions of the model, represented by a solid line. In this case, $E_K$ is calculated from equations \ref{eq:6} and \ref{eq:7} with the experimental cross section and this result is used to calculate the mobility \cite{borg90,borg88}.

The initial 3 orders-of-magnitude decrease of $\mu_0$ is well described up to $N=10.5$ atoms/nm$^3,$ where electron self-trapping begins \cite{borg90}. This result shows that the kinetic picture remains valid up to very high $N,$ if the gas-phase cross section is turned into an effective one by the multiple scattering effect.

\subsection{Localization at high density}
At very high $N$ in repulsive gases such as He and Ne, the kinetic description of the mobility fails, as shown in Figure 8, because a different physical phenomenon sets in: electron self-trapping \cite{hern}.

\begin{figure}\begin{center}
\includegraphics[scale=0.4]{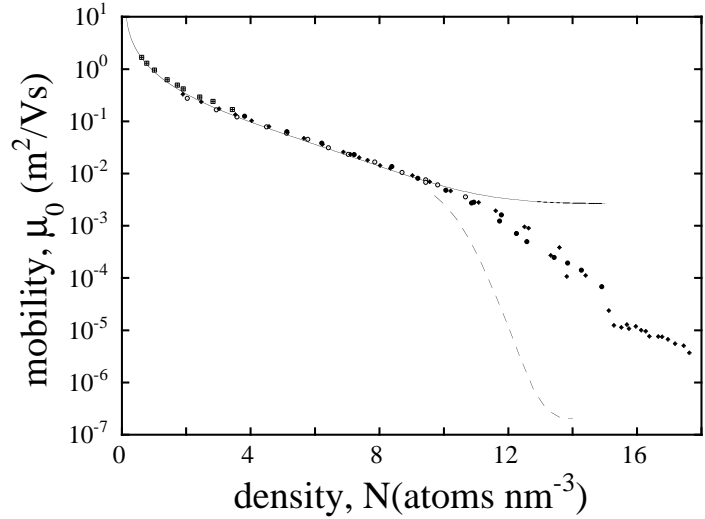}
\caption{\small  Zero-field mobility$ \mu_0$ in Ne at $T=45$ K \cite{borg90}. Solid line: heuristic model. Dashed line: localization model.}\end{center}
    \label{figure8}
\end{figure}

The interaction of an excess electron with the atoms of these gases is mainly repulsive and $V_0 (N)>0.$ The electrons may lower their free energy by trapping themselves in lower--than--average density fluctuations. If the local density $N_i$ in the fluctuation is small enough, the decrease $V_0(N)-V_0(N_i)$ compensates the increase of kinetic energy of the trapped electron due to localization and to the work spent to expand the cavity \cite{borg90,hern}. At constant $T,$ the excess free energy favors the localized state for $N$ greater than a given threshold, as can be observed in Figure 8, in which the mobility in Ne for $T=45.0$ K shows an enhanced slope around $N=10.5$ atoms/nm$^3$ cite{borg90}.

\begin{figure}\begin{center}
\includegraphics[scale=0.4]{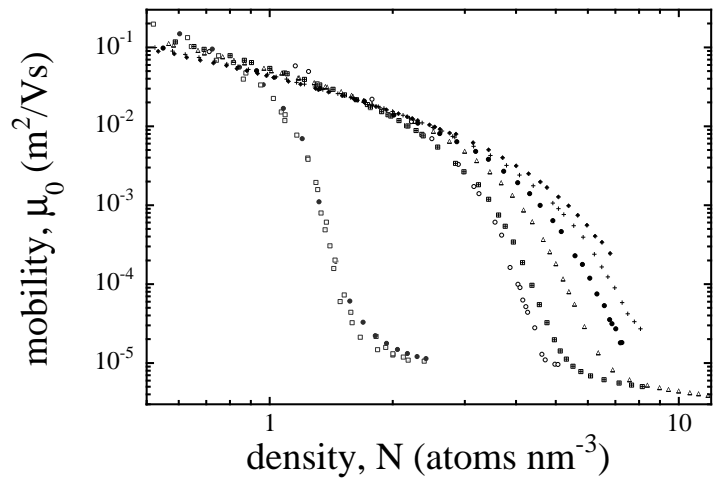}
\caption{\small  Mobility $\mu_0$ in He for $T=4.2$ \cite{levi}, $T=20.3$ \cite{jan}, $T=$26.1, 34.5, 45.0, 54.5, and 64.4 K \cite{borg02}  (from left to right). }\end{center}
    \label{figure9}
\end{figure}
The threshold density increases with increasing $T,$ mainly because of the large increase of the cavity expansion work as a consequence of the superlinear increase of pressure. This behavior is shown for He for several temperatures in Figure 9 \cite{borg02,levi,jan}. The localization-delocalization transition shifts to higher $N$ as $T$ is increased. 

Although the dynamics of bubble formation is not completely understood yet \cite{rosen,sakai}, the very simple quantum mechanical model of a particle in a spherically symmetric square box can grasp the essential physical features of localization \cite{borg02,borg90}.

The well depth is related to$ N$ by means of $ V_0 (N).$ The well is not completely empty because, for not too low $ T,$ the atoms have sufficient energy to penetrate into the cavity. Thus, the net well depth is actually $V_0(N)-V_0(N_i),$ where $N_i$ is the density of atoms inside the bubble. Basic Quantum Mechanics is used to solve for the energy eigenvalue of the ground state of the electron in the bubble, although more refined self-consistent model can be used \cite{hern}. The same approach has been used to interpret the results of infrared absorption of electron bubbles in liquid helium \cite{grimes}.

The free energy is readily calculated and the excess free energy is minimized with respect to bubble radius and filling fraction in order to give the most probable state.

\begin{figure}\begin{center}
\includegraphics[scale=0.4]{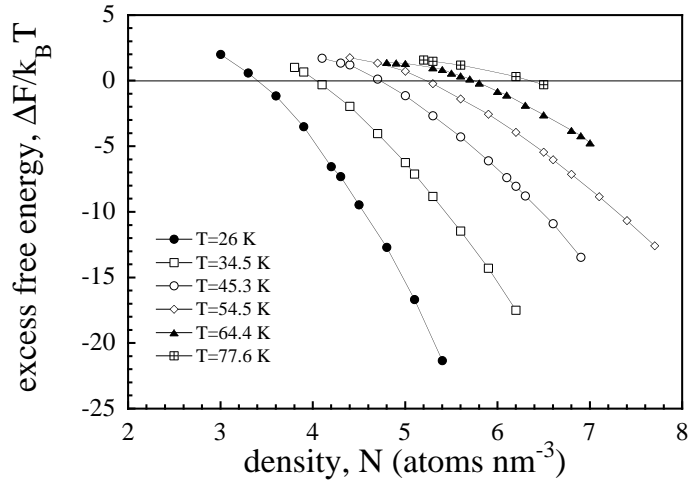}
\caption{\small Density and temperature dependence of the excess free energy of electrons in dense He gas  \cite{borg02}.  }\end{center}
    \label{figure10}
\end{figure}

In Figure 10  the computed minimum excess-free energy $\Delta F$ of electrons in dense Helium gas is shown as a function of density for several temperatures. We note that the density at which $\Delta F=0,$ i.e., the density at which the extended and localized states are equiprobable, shifts to larger values as the temperature is increased, in fair agreement with the expected behavior. Similar results are obtained also for Ne \cite{borg90}.

The mobility $\mu$ is then calculated as a weighted sum of that of the free electron, computed with the heuristic model
described previously, and that of the localized state that is given by standard hydrodynamics because of the large size and effective mass of the electron bubble. The equilibrium fraction of free to localized states is easily computed as $n_F/n_B = exp(-\Delta F/k_\mathrm{B}T),$ where $\Delta F$ is the minimum excess free energy. The dashed line in Figure 8 gives an example of the results.

This simple localization model describes qualitatively well the experimental observations, including, for instance, the shift of
the localization threshold with increasing $T.$ The agreement with the absolute mobility values is, however, quite poor, mainly
because the physical problem cannot be reduced to a two-species problem, the free and the optimum localized state. Probably, there is a whole distribution of bubbles of different radii, which differently contribute to the average mobility. However, the
basic physics of localization is reasonably well explained.

\begin{figure}\begin{center}
\includegraphics[scale=0.4]{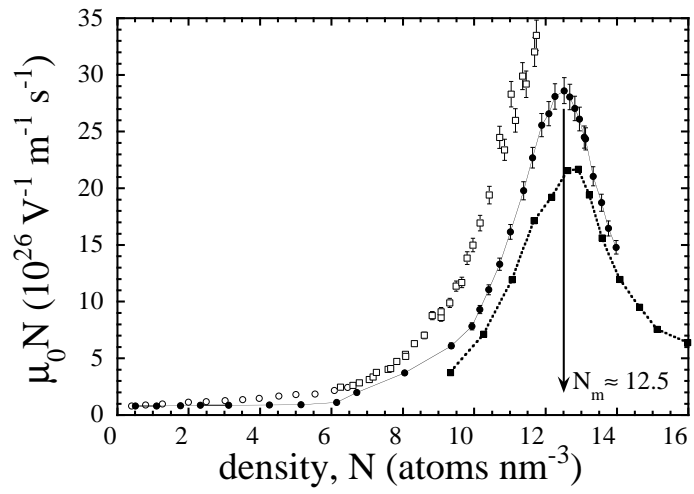}
\caption{\small Density dependence of the zero-field density-normalized mobility $\mu_0 N$ of electrons in dense Ar gas for $T=162.3-162.7$ K (open symbols) \cite{bsl} and for $T=151.5 K$ (closed dots) \cite{borg01}. The electron mobility in liquid Ar at coexistence is also shown (closed squares) \cite{busch}.}\end{center}
    \label{figure11}
\end{figure}

\subsection{Mobility maximum in very dense Argon} 
In an attractive gas, such as Ar, the situation at very high density is quite different with respect to the case of repulsive gases. Apparently, electron localization might occur in a larger-than-average density fluctuation, owing to the negative value of $V_0(N).$ As a matter of fact, however, this does not happen because, even at close packing, the excess free energy is not large and negative enough.

In spite of this, the behavior of the electron mobility in dense Ar is very interesting. In Figure 11 the zero-field density-normalized mobility $\mu_0 N$ is shown as a function of the density for $T=151.5$ K, just a few tenths of a degree above the critical temperature $T_c=150.9$ K \cite{borg01}.

The mobility in the dense gas shows a very pronounced peak at the very same density, at which a similar maximum is observed in liquid Ar at coexistence \cite{busch}. The peak in the gas can be somehow related to the RT minimum in the cross section \cite{borg01}, thus indicating that the RT minimum is plausibly present also in the liquid  \cite{christ}.

\section{Resonant electron attachment to O$2$}
Another interesting phenomenon, which is intimately related to the existence of a $N-$dependent contribution to the ground-state energy of an electron in a dense medium, is the resonant attachment of electrons to O$_2$ molecular impurities in a gas.

The study of attachment may be considered as a non-conventional way to carry out molecular spectroscopy in dense gases \cite{hern}. The electron attachment mechanism is the well-known Bloch--Bradbury one \cite{bb}. O$_2$ has an electron affinity of 0.46 eV. In a first step, an unstable negative ion is formed in a vibrational excited state, O$_{2}^{-*}.$ Then, the excess energy is carried away by a stabilizing collision with an atom of the host gas, as summarized in the following scheme
\begin{eqnarray}
\mathrm{O}_2 +e &\rightarrow &\mathrm{O}_2^{-\star} \nonumber \\
\mathrm{O}_2^{-\star}  + M & \rightarrow & \mathrm{O}_2^{-} +M \label{eq:12} 
\end{eqnarray}

\begin{figure}\begin{center}
\includegraphics[scale=0.4]{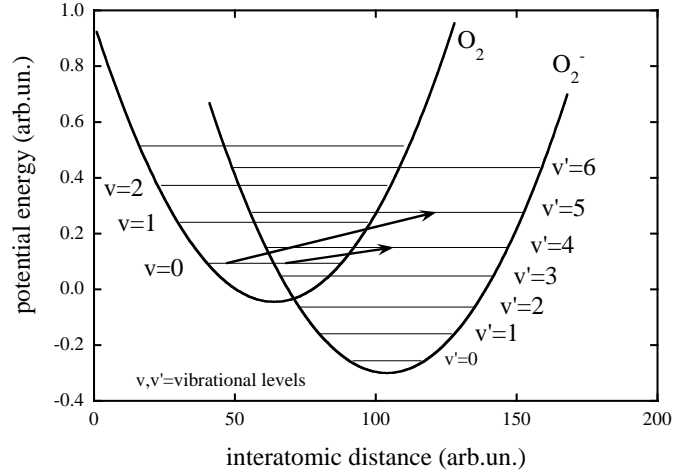}
\caption{\small Scheme of the energy curves involved in the formation of the negative molecular oxygen ion according to the Bloch-Bradbury mechanism. Arrows indicate the first two accessible excited vibrational levels of the ion.}\end{center}
    \label{figure12}
\end{figure}
where $M$ represents one atom of the host gas. A scheme of the energy curves involved in the process is reported in Figure 12.

In {\em vacuo}, the energies the energies of the two first accessible vibrational levels, namely those with $v'=4$ and $v'=5,$ respectively, on averaging over the spin-orbit split doublets, are $E_R^{(4)}=91$ and $E_R^{(5)}=207$ meV above the ground state of the neutral molecule.

\begin{figure}\begin{center}
\includegraphics[scale=0.4]{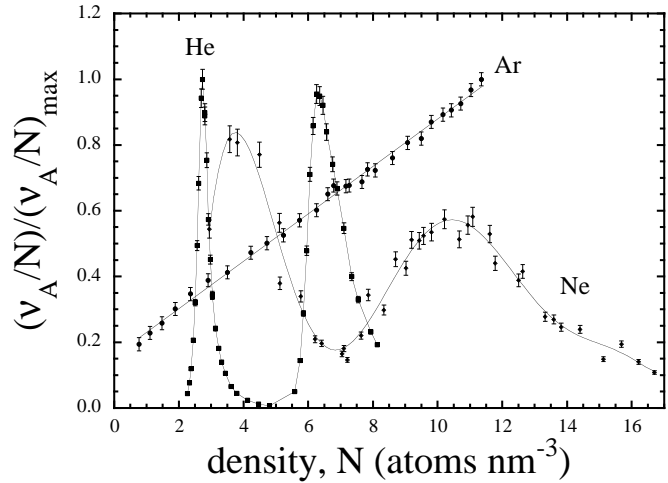}
\caption{\small Reduced attachment frequency $\nu_A /N$ for He at  $T=54.5 K,$ for Ne at $T=46.5$ K, and for Ar at $T=162$ K as a function of N \cite{neri}.}\end{center}
    \label{figure13}
\end{figure}

Once stabilized, ions are then detected in a traditional drift experiment. The current due to drifting electron decreases exponentially in time due to the formation of the very slow ions. An analysis of the current waveform gives electron lifetime, whose inverse is the attachment frequency $\nu_A.$ In Figure 13 the reduced attachment frequency $\nu_A /N$ is shown as a function of $N$ for He, Ne, and Ar [34]. Strong peaks appear for He and Ne, while no structure is present in Ar.

It can be shown that the density-normalized attachment frequency is  $\nu_A /N \propto F(E_R),$ where $F(E_R)$ is the electron energy distribution function evaluated at the resonance energy \cite{bruschi}. If the excess electron ground state energy were unaffected by density, $\nu_A/N$ should be independent of $N.$ The existence of attachment peaks in He and Ne for a density $N=N_R$ is easily explained by noting that $V_0(N)>0$ for these gases. An increase of $N$ shifts the bottom of the electron energy distribution function that is sampled by the attachment process at the fixed energy $E_R.$ This process is schematically depicted in Figure 14.

Actually, in a dense gas, whose electron-atom interaction is repulsive, a partially empty cavity is formed around the ion and the condition for energy conservation at resonance must include the excess free energy spent to create the bubble. In any case, at the density $N_R$ of the peak the mean electron energy equals the resonance energy. The peaks are thus a sort of replica of$ F(\epsilon).$

At even higher $N,$ the shift in energy of the distribution becomes eventually so large that attachment to the second accessible vibrational level of O$_2^-$ can take place and a second peak appears \cite{neri}.

In Ar no such structures exist because the polarization contribution to $V_0(N)$ is so large as to make it negative and because the polarization energy of the ion cannot be neglected. In this case the average electron energy is shifted downwards, in the direction of the lower lying vibrational levels of the molecular ion ($v'=2,$ for instance), but these levels are also shifted downwards by nearly the same amount because of the ion polarization energy in such a way that the condition for resonance is never met. 
\begin{figure}\begin{center}
\includegraphics[scale=0.4]{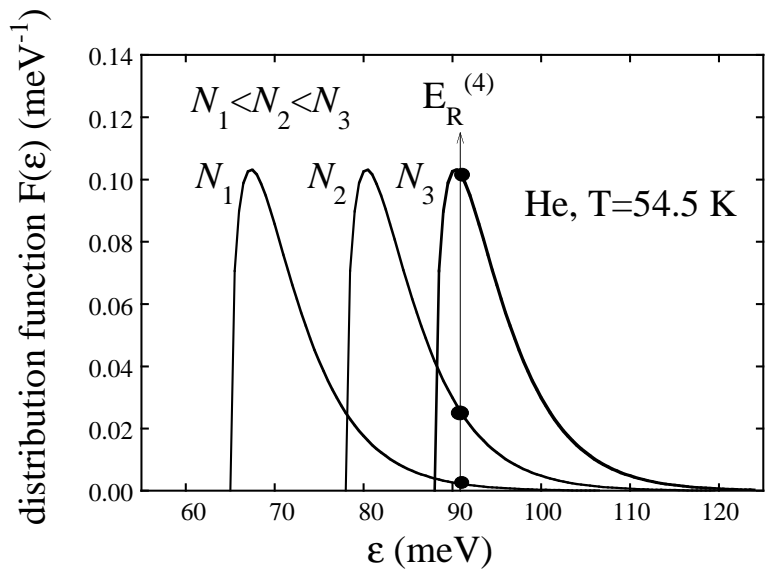}
\caption{\small Shifted Maxwellian distributions in He at $T=54.5$ K \cite{neri}. The resonant energy is  91 meV. Dots show where the distribution function is sampled in the resonant attachment process.}\end{center}
    \label{figure14}
\end{figure}

It has to be noted, that, at variance with the scattering processes involved in the mobility experiments, in which the electron kinetic energy is shifted only by the positive amount $E_K(N)$ both in repulsive as well as in attractive gases, the total energy shift $V_0(N)$ must be taken into account for energy conservation in the resonant electron attachment process

\section{O$_2^-$ Transport at high density}
The negative O$_2^-$ molecular ion formed by electron attachment has a complex structure because of both electrostriction and quantum exchange forces. An electrostriction--induced solvation shell surrounds the ion \cite{atk}. Owing to the symmetry of the polarization potential, the ion-solvation shell complex can be assumed to be a sphere with a several  large radius.

It easy to show that the local density profile N(r) is given by \cite{atk,cpl}
\begin{equation}
-V(r) = K^2\left(N\left(r\right)\right)\int\limits_N^{N(r)} \frac{1}{N'}\left( 
\frac{\partial P\left( N'\right)}{\partial N'}\right)_T \mathrm{d} N'
\label{eq:elstr}
\end{equation}
where $V(r)$ is the ion-atom interaction potential, $ K$ is the dielectric constant of the gas and $N$ and $P$ are its density and pressure, respectively. A typical density profile surrounding the negative oxygen ions in Ar for $T=151.5$ K is shown in Figure 15.
\begin{figure}
\begin{center}
\includegraphics[scale=0.4]{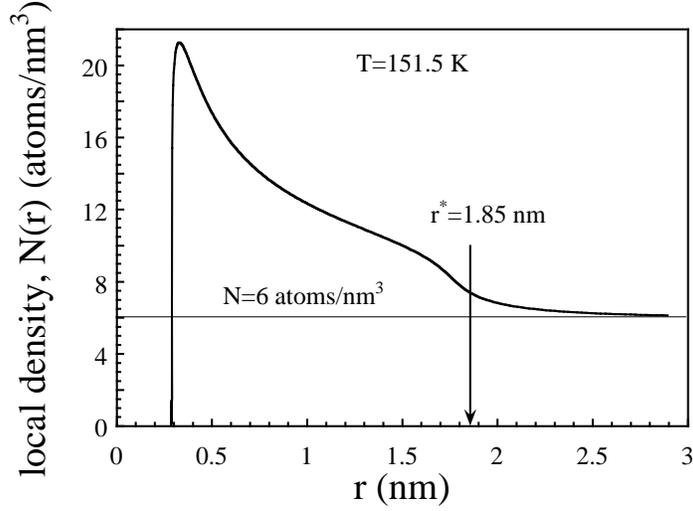}
\caption{\small Electrostriction--induced density profile around an O$_2^-$ ion in Ar gas for $T=151.5$ K. The density of the unperturbed fluid far from the ion is $N=6$ atoms/nm$^3$ \cite{cpl}. $r^\star$ is explained in the text.}\end{center}
    \label{figure15}
\end{figure}

Owing to its large size, the ion can be thus used as a probe to investigate the transition between the hydrodynamic transport regime, typical of the liquid, and the kinetic regime, typical of the dilute gas.
\begin{figure}
\begin{center}
\includegraphics[scale=0.4]{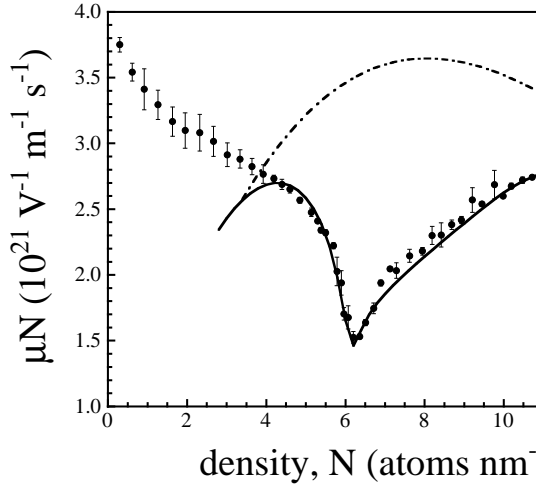}
\caption{\small Reduced O$_2^-$ mobility in Ar at $T=151.5$ K \cite{cpl}. Dash--dotted line: Stokes mobility. Solid line: Stokes formula modified by electrostriction and criticality.}\end{center}
    \label{figure16}
\end{figure}

In Figure 6 the measured zero-field density-normalized mobility of the O$_2^-$ ion in Ar gas at $T=151.5$ K is shown \cite{cpl}. The critical parameters of Ar are $ T_c= 150.86$ K and $N_c=8.08$ atoms nm$^{-3}.$ The experimental data show a deep drop of $\mu N$ at $N=6.25$ atoms/nm$^3 < N_c.$

Classical hydrodynamics predicts that m is related to the gas viscosity $\eta$  by means of the Stokes formula
\begin{equation}
\mu =\frac{e}{6\pi\eta R}
\label{eq:stks}
\end{equation}
where $R$ is the ion effective hydrodynamic radius. The dash--dotted line in Figure 16  represents the prediction of the Stokes formula with a reasonable choice of the ion radius, $R=0.58$ nm \cite{cpl}. Only at the highest $N,$ equation \ref{eq:stks} approaches the experimental data but it does not reproduce at all the experiment for $N\approx N_c.$ Similar behavior is observed also in Ne \cite{borg93}.

The hydrodynamic prediction can be reconciled with the data if electrostriction in a near-critical fluid is considered. Electrostriction induces a local density and viscosity enhancement around the ion. Criticality makes a local long-range order to appear in the gas. A thick layer of strongly correlated fluid is this dragged along by the ion. In this way the effective hydrodynamic ion radius depends on how close criticality is approached.

Near the critical point, the correlation length $ \xi$  is the longest and the effective ion radius is the largest. Because of electrostriction, however, criticality in the fluid surrounding the ion is attained at a given distance from the ion only when the unperturbed fluid at large distance is below the critical density. This explains why the mobility drop occurs for $N<N_c.$

A simple model has been developed to account for electrostriction and criticality \cite{cpl}. The effective ion hydrodynamic radius in \ref{eq:stks} is assumed to be given by
\begin{equation}
R_e =b_0 +b_1N + b_2\xi\left(N\left(r\right)\right)
\label{eq:re}
\end{equation}
where $b_0,$ $b_1,$ and $b_2$ are adjustable parameters. $\xi\left(N\left(r\right)\right)$  is the local value of the correlation length of the fluid.  $\xi(N(r)\propto \sqrt{S(N(r))}$ \cite{stanley}, where $S(N(r))$ is the long wavelength limit of the static structure factor evaluated as a function of the local density at a given distance $r$ from the ion. The linear contribution $b_0+b_1N$ interpolates between the radii of the first and second solvation shells, which appear at high $N$ even far from the critical point.

The distance, at which the static structure factor is to be evaluated, must be chosen so as to reproduce the experimental mobility minimum. If $r=r^\star =1.85$ nm is chosen, and if also the local value of the viscosity is used in equation \ref{eq:stks}, the electrostriction-modified Stokes formula agrees very nicely with the data, as shown by the solid line in Figure 16.

For $N<4$ atoms nm$^{-3},$ in any case, hydrodynamics is no longer valid and the modified Stokes formula fails to reproduce the data, even though $N$ is still too large for kinetic theory to apply. 

\section{Red--shift of infrared excimer fluorescence in a dense gas}

The unifying concept of the previously reported phenomena is the $N-$de\-pen\-dent shift of the ground state energy of a free electron in a dense gas. The same concept explains the experimental results in a different situation, in which the electron is not free but it is bound to a molecule.

In a noble gas, such as Xe, excited by means of energetic electrons, transient excited molecular species or excimers, such as Xe$_2,$ are formed. The ground state of this molecule is dissociative but higher lying states may be bound \cite{mul}.

In the deexcitation path leading again to two neutral atoms, the transition from the first excited level of the excimer to its ground state is the most energetic and leads typically to vacuum ultraviolet (VUV) fluorescence. In Xe, this occurs at 173 nm and is exploited to detect ionizing particles. In the context of high-energy physics, fluorescence is termed scintillation \cite{kn}.

A broad band of near infrared (NIR) fluorescence of Xe$_2$ in pure Xe gas and in Ar-Xe gas mixture has been recently discovered \cite{ian}. It occurs around $7.9\times 10^5$ m$^{-1},$ corresponding to a transition energy of nearly 1 eV. The fluorescence thus involves a less energetic transition between higher lying excimer levels.

The most important feature of this NIR band is that its center is red-shifted if the gas density is increased. The shift in pure Xe is larger than in the 90\% Ar--10\% Xe mixture, as shown in Figure 17.

The experimental results are easily explained if the excimer is modeled as consisting of an ionic Xe$_2^+$ core plus a delocalized electron in a Rydberg--like state of very large radius. The electron interacts with many atoms of the host gas simultaneously.

As a consequence, its energy is shifted by $V_0(N),$ as in the case of the free excess electron. Owing to the relatively small value of $N,$ Fermi's formula can be used \cite{fermi}
\begin{equation}
V_0 (N) =\frac{2\pi\hbar^2}{m}Na
\label{eq:fermi}
\end{equation}
where $a$ is the electron--atom scattering length. 

Moreover, the orbit encompasses many atoms of the gas, which screen the Coulomb interaction between the electron and the ionic core. This effect is a solvation effect that always redshifts the wavelengths emitted in the transitions.
\begin{figure}
\begin{center}
\includegraphics[scale=0.4]{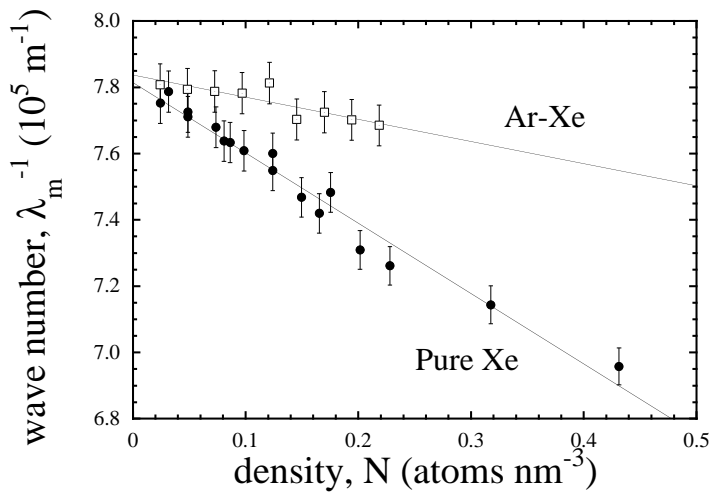}
\caption{\small Red--shift of the center of the NIR fluorescence band of Xe$_2$ excimers in pure Xe and in an Ar--Xe mixture \cite{ian}. The solid lines are equation \ref{eq:exc}.}\end{center}
    \label{figure17}
\end{figure}
It can be shown that, for not too large N, the central wave number of the NIR fluorescence can be written as \cite{ian}
\begin{equation}
\lambda_m^{-1} = \lambda_{m,0}^{-1}-\left(  \lambda_{m,0}^{-1}\frac{2\alpha}{\epsilon_0} -\frac{\hbar}{mc}a\right) N
\label{eq:exc}
\end{equation}
where $\lambda_{m,0}^{-1}$ is the wave number of the center of the band in the zero--density limit and a is the atomic polarizability of the gas. The first term in brackets is the solvation contribution that always produces a red--shift of the band. 

The second term is due to the shift of the ground state energy of the electron in a dense medium because of its quantum nature. Depending on whether $a$ is positive or negative, it reduces or enhances the red--shift produced by solvation. In the case of Xe and Ar, $a<0.$ In the mixture, the Xe$_2$ excimers are surrounded, on average, by Ar atoms, so that the parameters $a$ and $\alpha$ of Ar have to be used, which yield a slope smaller than for pure Xe. 

Equation \ref{eq:exc} predicts a linear dependence of the shift of the center of the band, as experimentally observed. In Figure 17 the lines through the data point are calculated by using this equation. The agreement is very good \cite{ian}].

\section{Conclusions}
The investigation of the transport properties of excess electrons in dense non-polar gases has produced a large wealth of data that gives a unified picture of the electron states and dynamics in a dense host.

At high densities, quantum effects of the electron--atom interaction in a dense environment come into play. The most important phenomenon is that the ground state energy of the excess electrons do now depend on the density of the gas. This effect modifies all the dynamic properties of electrons. In particular, the gas--phase cross section is progressively turned into an effective, density-dependent one as the density is increased. This is the way, along which the conduction properties of dense liquids might be bridged to the properties of dilute gas.

It is interesting to realize that the results of the analysis of the transport properties shed light also on apparently uncorrelated phenomena, such as excimer fluorescence in a dense gas. It would be interesting to see if this concept could be helpful in understanding other phenomena, such as electron--ion recombination in a dense gas.

It is clear that these results have been obtained for noble gases, which are the simplest possible systems, and, under this respect, they are paradigmatic. An interesting future goal might be to see if measurements in more complex systems, such as molecular gases, for instance, may fit into the unified scheme emerging from noble gases.

\newpage 
\bibliography{transport}
 \end{document}